\documentclass[aps,twocolumn,showpacs,superscriptaddress]{revtex4}
\usepackage{graphicx}

\begin{document}

%Version 1 Gennady Nov. 11/03
%Version 2 Fausto  Nov 13/03 add figures and correct few misprints
\title{A Model for Quantum Jumps in Magnetic Resonance Force Microscopy} 
\author{G.P. Berman}
\affiliation{Theoretical Division and CNLS, MS B213, 
Los Alamos National Laboratory, Los Alamos, NM 87545} 

\author{F. Borgonovi}
\affiliation{Dipartimento di Matematica e Fisica, Universita Cattolica, 
via Musei 41, 25121 Brescia, Italy}
\affiliation{INFM, Unit\`a di Brescia and INFN, Sezione di Pavia, Italy}

\author{V.I. Tsifrinovich}
\affiliation{IDS Department, Polytechnic University, Six Metrotech
Center, Brooklyn, New York 11201}

\date{\today} 

\begin{abstract} 

{We propose a simple model which describes the statistical properties of 
quantum jumps in a single-spin measurement using the oscillating cantilever-driven 
adiabatic reversals technique in magnetic resonance force microscopy. 
Our computer simulations based on this model predict the average time interval 
between two consecutive quantum jumps and the correlation time
to be proportional to the characteristic time of the magnetic noise 
 and inversely proportional to the square of 
the magnetic noise amplitude.}

\end{abstract} 

\maketitle 

\section{Introduction}

Recent achievements in magnetic resonance force microscopy (MRFM) promise a single 
spin detection in the near future \cite{1}. One may expect that single spin signal 
will represent a random sequence of quantum jumps. The important problem for the 
theory is modeling of quantum jumps in MRFM and the computation of their statistical
 characteristics. In this paper we propose a simple model which describes quantum 
jumps in MRFM single spin detection. We consider the oscillating cantilever-driven 
adiabatic reversals technique (OSCAR) which currently is the most promising approach 
for single spin detection \cite{1}. In OSCAR the cantilever tip (CT) with a 
ferromagnetic particle oscillates, causing the periodic adiabatic reversals of the 
effective magnetic field on spin. The spin follows the effective magnetic field 
causing a tiny frequency shift of the CT vibrations which is measured.

In the second section we consider the Schr\"odinger dynamics of the CT-spin system 
which underlines our model of quantum jumps. In the third section we describe our 
model and present the results of the computer simulations. In Conclusion we discuss 
our results.

\section{Schr\"odinger dynamics of the CT-spin system}

\begin{figure}
\includegraphics[scale=0.44]{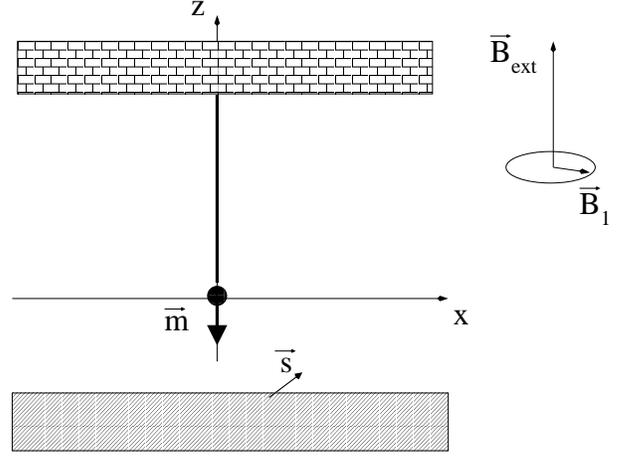}
\caption{
Single spin OSCAR MRFM setup. ${\vec B}_{ext}$ 
is the external permanent magnetic field; $B_1$ is the rotating {\it rf} field; 
${\vec m}$ is the magnetic moment of the ferromagnetic particle; 
${\vec s}$ is a spin near the surface of the sample.}
\label{f1}
\end{figure}

We consider a vertical cantilever with a ferromagnetic particle attached to the 
CT and oscillating along the $x$-axis which is parallel to the surface of the sample. 
(See Fig. \ref{f1}). 
The Hamiltonian of the CT-spin system in the system of coordinates which rotates 
with a circularly polarized {\it rf} field can be written as
$$
{\cal H}={{1}\over{2}}(p_c^2+x^2_c)+\varepsilon 
S_x+2\eta x_cS_z+\Delta(\tau) S_z.\eqno(1)
$$
Here $p_c$ is the effective momentum of the CT, $x_c$ is 
its coordinate, the first term describes the effective 
energy of the CT; the second term describes the interaction between the 
spin and the {\it rf} field; the third term describes the interaction 
between the CT and the spin; and the last term describes the 
interaction between 
the spin and the random magnetic field which causes a deviation of spin 
from the effective magnetic field when the latter passes through 
the transverse $x-y$-plane \cite{2,3}. 
All quantities in (1) are dimensionless: $p_c$ and $x_c$ are expressed in terms of 
the ``quantum units'' $X_0$ and $P_0$
$$
X_0=(\hbar\omega_c/k_c)^{1/2},~P_0=\hbar/X_0.\eqno(2)
$$
The dimensionless parameters $\varepsilon$, $\eta$, and $\Delta(\tau)$ are 
defined as
$$
\varepsilon=\gamma B_1/\omega_c,~\eta=\gamma(\hbar/k_c\omega_c)^{1/2}|\partial 
B_z(x)/\partial x|/2,\eqno(3)
$$
$$
\Delta(\tau)=\gamma\Delta 
  B_z(\tau)/\omega_c,~\tau=\omega_c t.
$$
Here $B_z$ is the $z$-component of the ``regular magnetic field'' which 
includes ${\vec B}_{ext}$ and the regular dipole-magnetic field produced 
by the ferromagnetic particle at the location of the spin; $\Delta B_z(\tau)$
 is the $z$-component of the random field with zero average value; $\gamma$ 
is the magnitude of the spin gyromagnetic ratio: $k_c$ and $\omega_c$ are 
the effective spring constant and fundamental frequency of the CT; and 
$\tau$ is the dimensionless time.

Using the parameters presented in \cite{1}
$$
\omega_c/2\pi=6.6 kHz,~k_c=6\times 10^{-4} N/m,\eqno(4)
$$
$$
 B_1=0.3 mT,~|\partial B_z(x)/\partial x|\approx 4.3\times 10^5 T/m,
$$
we can estimate all parameters in (1) except of $\Delta (\tau)$:
$$
X_0=85 fm,~P_0=1.2\times 10^{-21} Ns,~\eta=0.078,~\varepsilon=1270.\eqno(5)
$$
To simplify computer simulations we considered the function $\Delta(\tau)$ 
to be a random telegraph signal with two values $\pm\Delta$. The time 
interval between two consecutive ``kicks'' of $\Delta(\tau)$ was taken 
randomly from the interval $(\tau_0-\delta \tau, \tau_0+\delta \tau )$, with 
the average time  interval, $\tau_0$, close  to the Rabi period $\tau_R$
$$
\tau_R=\omega_c(2\pi/\gamma B_1)=2\pi/\varepsilon=4.95\times 10^{-3}.\eqno(6)
$$
We choose the initial state of the CT to be a coherent 
quasiclassical state, and the average spin to be pointed along 
the ``regular'' effective magnetic field with the components 
$(\varepsilon, 0, 2\eta x_c)$. 

Below we describe briefly the results of our computer simulations. Our 
simulations reveal the formation of a Schr\"odinger cat state for the CT: 
the probability distribution function 
$P(x,\tau)\equiv\Psi^\dagger(x,\tau)\Psi(x,\tau)$ 
splits into two peaks. Similar to our previous computations \cite{4,5,6} 
one peak corresponds to the average spin pointing in the direction of the 
effective magnetic field, while the other one corresponds to the opposite 
direction of the average spin. The two peaks oscillate with slightly different
 periods due to the back action of the spin as  expected in the OSCAR technique 
\cite{7,8}. Unlike our previous papers \cite{4,5,6}, the appearance of two peaks 
is not connected with the initial deviation of the spin from the direction of
 the effective field. It is induced by the action of the random field 
$\Delta(\tau)$ which causes this deviation in the process of the spin-CT dynamics.
 It was shown in \cite{5,6} that the interaction with the environment quickly
 destroys quantum coherence between the two peaks: the Schr\"odinger cat state 
of the CT quickly transforms into a statistical mixture of the two possible CT 
trajectories. Thus, the random magnetic field $\Delta(\tau)$ causes the quantum 
jumps: spin may ``jump'' to the opposite direction relative to the effective 
magnetic field and, correspondingly, the CT may slightly change the period of 
its oscillations.

Unfortunately, our present simulations consume too much computer time to be able 
to reveal more than one spin jump. This approach, clearly, does allow one not  
to study statistical characteristics of quantum jumps. Consequently we 
developed a simplified model which describes statistical properties of the spin jumps. 

\section{Simple model of quantum jumps}

\begin{figure}
\includegraphics[scale=0.33]{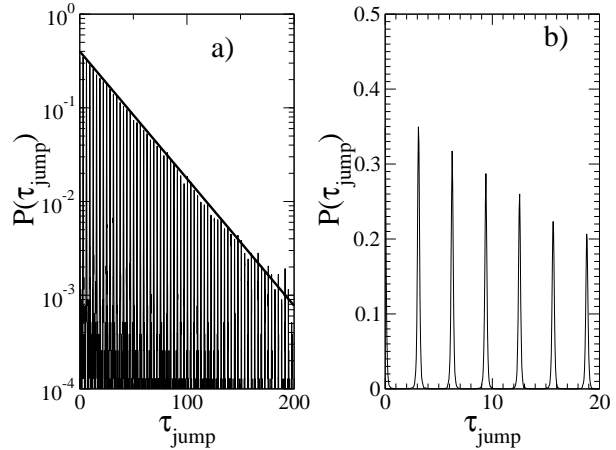}
\caption{
(a) Distribution function of time intervals between two consecutive 
quantum jumps for $\Delta=100$ $\tau_0=0.01$ and $10^9$ kicks;  
the solid line is a fit with $\tau_d=32$; 
(b) Enlargement of (a)}
\label{f2}
\end{figure}

We assume that every ``kick'' provided by the function $\Delta(\tau)$ is followed 
by the collapse of the wave function. Before the kick, the spin points in (or opposite 
to) the direction of the effective magnetic field. After the kick there appears the 
finite angle $(\Delta\Theta)$ between the new direction of the effective field and 
the average spin. The probability for the spin to ``accept''
the ``before-kick'' direction relative to the new effective field is $\cos^2(\Delta\Theta/2)$.
 The probability to ``accept'' the opposite direction, i.e. the probability of a 
quantum jump is $\sin^2(\Delta\Theta/2)$. (It is clear that a significant probability 
of a quantum jump appears only when the effective field passes the transversal $x-y$-plane. 
In the transversal plane, the effective field has the minimum value.
 Thus, after every kick of the random field our computer code 
decides the ``fate'' of the spin in accordance with the 
probabilities of two events: to restore the previous direction relatively 
to the effective field, or to experience a quantum jump. 
In our model the CT experiences harmonic oscillations
$$
x^{(\pm)}_c(\tau)=x_m\cos(1\pm\delta\omega)\tau,\eqno(7)
$$
where $(\pm)$ correspond to two CT trajectories with the spin pointing in the 
direction of (or opposite to) the corresponding effective field, and 
$\delta\omega$ is the CT frequency shift. The components of the effective field 
are given by
$$
(\varepsilon,0,2\eta x_c(\tau)+\Delta(\tau)).\eqno(8)
$$
From the experimental data \cite{1} for the CT amplitude $X_m=10$ nm we obtain 
$x_m=1.2\times 10^5$. The frequency shift, $\delta\omega$, can be estimated as \cite{8}
$$
\delta\omega=\Delta\omega/\omega_c={{2G\mu_B}\over{\pi X_mk_c}}=4.2\times 10^{-7}.\eqno(9)
$$

Note that our model contains two important simplifications: first,
we assume that the wave function collapse occurs immediately
after the ``kick'' of the random field. Thus, we ignore the finite 
time when the spin-CT system is in an entangled state.
Second, in a real situation the deviation of the spin from the
effective field is a ``quasi--resonance'' process caused by the cantilever 
modes whose frequencies are close to the Rabi frequency. In our model this 
deviation appears as a result of the ``kick'' of the random field.

\begin{figure}
\includegraphics[scale=0.35]{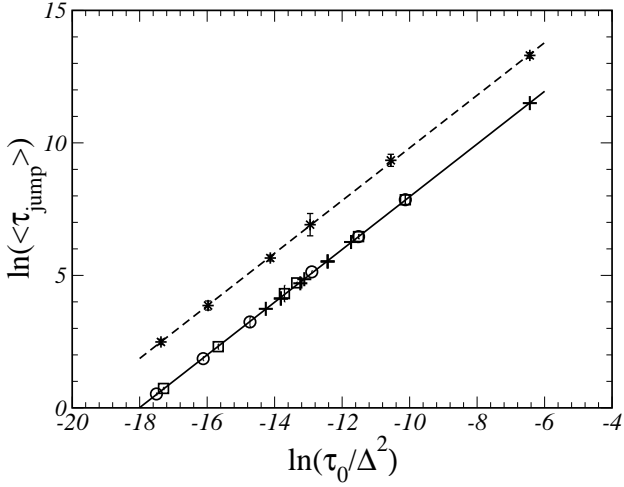}
\caption{
Dependence of the average time intervals between
two consecutive quantum jumps on $\tau_0/\Delta^2$.
The full line corresponds to the value $x_m=1.2 \times 10^{-5}$,
the squares represent $\delta\omega = 4.2 \times 10^{-8}$,
the crosses $\delta\omega = 4.2 \times 10^{-7}$,
the circles $\delta\omega = 4.2 \times 10^{-6}$
The dashed line corresponds to the value $x_m=7.2 \times 10^{-5}$.
Data  have been obtained by varying parameters
$\Delta$ and $\tau_0$ in the ranges:
$10<\Delta<300$ and $0.001<\tau_0< 1$.
}
\label{f3}
\end{figure}

Below we describe the results of our simulations. 
Fig. \ref{f2}  demonstrates a typical probability distribution of time intervals, 
$\tau_{jump}$, between two consecutive quantum jumps. The probability distribution 
is a sequence of sharp peaks at $\tau_{jump}=\tau_{n}= n\pi$ with the Poisson-like 
amplitude
$$
P(\tau_{n})\sim\exp(-\tau_{n}/\tau_d).\eqno(10)
$$
(Certainly, $P(\tau_{jump})=0$ at $\tau<\tau_0-\delta\tau$.) The sharp peak 
appears as the probability of the quantum jump is significant when the spin 
passes through the transversal plane, i.e. every half-period of the CT 
oscillation which is equal to $\pi$. The average value of the time interval 
$\langle\tau_{jump}\rangle$ was found to be
$$
\langle\tau_{jump}\rangle\approx\tau_d,\eqno(11)
$$
with an error less than 3\%. The standard deviation is equal to $\tau_d$ 
with the same accuracy
$$
(\langle\tau^2_{jump}\rangle-\langle\tau_{jump}\rangle^2)^{1/2}\approx \tau_d.\eqno(12)
$$

We studied the dependence of the average 
value $\langle\tau_{jump}\rangle$ on the parameters of our model.
We have found that   $\langle\tau_{jump}\rangle$ 
does not depend on $\delta\tau$ or $\delta \omega$. 
(We varied  $\delta\tau$ from $0$ to $\tau_0$ and changed 
$\delta\omega$ up to one order of magnitude.)
At a fixed value of the amplitude
$x_m$ the value of  $\langle\tau_{jump}\rangle$
 is approximately proportional to $\tau_0/\Delta^2$. 
Fig. \ref{f3}
 demonstrates this dependence. 

The best fit for the numerical points in Fig. \ref{f3} is given by
$$
\ln\langle\tau_{jump}\rangle=p +q \ln(\tau_0/\Delta^2).\eqno(13)
$$
For $x_m=1.2 \times 10^5$ we have 
$p=17.9$, $q=0.993$. For the six fold value $x_m = 7.2 \times 10^5$
we obtained the same value of $q$ and $p=19.743$.
If we estimate the amplitude of the random CT vibrations near the 
Rabi frequency 
as 1 pm, then $\Delta=1.8.$ Putting $\tau_0 = \tau_R$ 
and the experimental value for $\tau_R$ (6) we obtain 
$\omega_c\langle\tau_{jump}\rangle=2.3$ s  for $x_m=10 nm$ and
$\omega_c\langle\tau_{jump}\rangle=14.5$ s  for $x_m=60 nm$.
These values are close to 
the experimental characteristic times of $3$ s and $20$ s
reported in \cite{1}.

Next we computed the correlation function for the CT frequency
shift
$$
C(\tau) = {
{\langle (\delta\omega(t)-\overline{\delta\omega})
(\delta\omega(t+\tau)-\overline{\delta\omega})\rangle }
\over
{\langle (\delta\omega(t))^2\rangle - {\overline{\delta\omega}}^2}
},
\eqno(14)
$$
where $\overline{\delta\omega}=\langle \delta\omega(t)\rangle=0$, 
and $\langle ...\rangle$ indicates an average over time.

\begin{figure}
\includegraphics[scale=0.35]{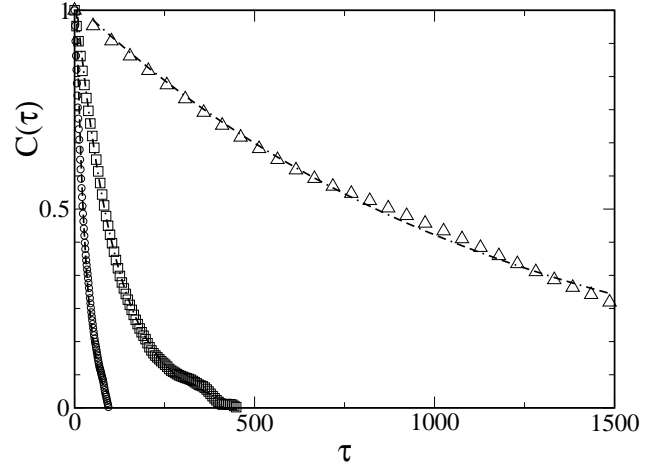}
\caption{
Correlation function of the CT frequency shift $C(\tau)$ for
different parameters: circles $\Delta=100, \tau_0=10^{-2}$,
squares $\Delta=50, \tau_0=10^{-2}$, triangles $\Delta=50, \tau_0=10^{-1}$.
In all cases
$\delta \tau = \tau_0/4$.
The dashed curves show the exponential approximation of the correlation
function $\exp(-\tau/\tau_c)$ with $\tau_c=23.91, 95.81, 1179.15  $, 
respectively.
}
\end{figure}

In Fig.4 we show the correlation function $C(\tau)$ for three different
values of parameters as indicated in the legend. As one can see, the 
general behavior is well described by the exponential function
(indicated by dashed  lines in Fig.4)  $\exp(-\tau/\tau_c)$.
The relation between the correlation time $\tau_c$ and $\langle\tau_{jump}\rangle$
was found to be $\langle\tau_{jump}\rangle \simeq 2.5 \tau_c$.

\section{Conclusion}

We developed a simple model which describes quantum jumps of 
spins in the OSCAR MRFM technique and which allowed us to 
compute the statistical characteristics of quantum jumps. 
In our model the average time interval $\langle\tau_{jump}\rangle$ 
between two consecutive jumps and the correlation time $\tau_c$ 
are proportional to the characteristic time of the magnetic noise,
and inversely proportional  to the square of the magnetic noise amplitude.
Using experimental parameters \cite{1} and a reasonable value 
for the CT  noisy amplitude we obtained the value of 
$\langle\tau_{jump}\rangle$ which is close to the experimental 
value of the characteristic time of OSCAR MRFM signal. 

\section*{Acknowledgments}

This work was supported by the Department of Energy (DOE) under
Contract No. W-7405-ENG-36, by the Defense Advanced Research Projects Agency 
(DARPA), by the National Security Agency (NSA),
and by the Advanced Research and Development Activity (ARDA).


\begin{thebibliography}{99} 
%
\bibitem{1}
H.J. Mamin, R. Budakian, B.W. Chui, D. Rugar, Phys. Rev. Lett. {\bf 91}, 207604 
(2003).
%
\bibitem{2}
D. Mozyrsky, I. Martin, D. Pelekhov, P.C. Hammel, Appl. Phys. Lett. {\bf 82}, 1278 
(2003).
%
\bibitem{3}
G.P. Berman, V.N. Gorshkov, D. Rugar, V.I. Tsifrinovich, Phys. Rev. B {\bf 68}, 
094402 (2003). 
%
\bibitem{4}
G.P. Berman, F. Borgonovi, G. Chapline, S.A. Gurvitz, P.C. Hammel, D.V. Pelekhov, 
A. Suter, V.I. Tsifrinovich, J. Phys. A: Math. Gen. {\bf 36}, 4417 (2003). 
%
\bibitem{5}
G.P. Berman, F. Borgonovi, H.S. Goan, S.A. Gurvitz, V.I. Tsifrinovich, Phys. Rev. B 
{\bf 67}, 094425 (2003).
%
\bibitem{6}
G.P. Berman, F. Borgonovi, V.I. Tsifrinovich, quant-ph/0306107 (2003).
%
\bibitem{7}
B.C. Stipe, H.J. Mamin, C.S. Yannoni, T.D. Stowe, T.W. Kenny, D. Rugar, Phys. Rev. 
Lett. {\bf 87}, 277602 (2001).
%
\bibitem{8}
G.P. Berman, D.I. Kamenev, V.I. Tsifrinovich, Phys. Rev. A {\bf 66}, 
023405 (2002).
              

               
\end{thebibliography}
\end{document}